\documentclass[11pt,a4paper]{article}
\usepackage{jinstpub}

\usepackage{booktabs}
\usepackage{siunitx}
\usepackage{xcolor}
\usepackage{graphicx}

\title{A Modular Datalogger and Slow-Control Platform for Physics Experiments with Time-Series Telemetry and Web Dashboards}

\author[a,1]{D.~Tagnani\note{Corresponding author.}}
\author[b]{N.~Toniolo}
\author[b]{B.~Gongora Servin}
\author[b]{T.~Marchi}
\author[b]{A.~Goasduff}

\affiliation[a]{INFN, Sezione di Roma Tre, Via della Vasca Navale 84, I-00146 Rome, Italy}
\affiliation[b]{INFN, Laboratori Nazionali di Legnaro, Viale dell’Università 2, 35020 Legnaro, Italy}

\emailAdd{diego.tagnani@roma3.infn.it}

\abstract{
We present a NIM 2U slow-control system designed for continuous monitoring of environmental and facility parameters during data taking in nuclear-physics experiments.
The architecture integrates a base controller board with eight channels for PT100/PT1000 probes and up to three plug-in extension boards, each providing SMA connections on the front panel.
All boards adopt a 16-bit SAR ADC (AD7689) with an external \SI{2.5}{V} reference, yielding an LSB of \SI{38.15}{\micro V} and ensuring metrological uniformity across heterogeneous channel types \cite{AD7689,REF192}.
The operating range for PT100/PT1000 probes spans \SI{-200}{\celsius} $\rightarrow$ \SI{+300}{\celsius}, with a design accuracy at the \SI{1}{\%} level.
The system includes modules dedicated to standard industrial signals 4--20~mA and 0--10~V (manual per-channel selection) and a module for current measurement in the \SI{10}{nA}--\SI{150}{\micro A} range with automatic gain-range selection.
The software chain publishes time series to Graphite and provides visualization through Grafana over Ethernet \cite{GraphiteOverview,GrafanaIntro}.
Representative performance measurements are reported for the RTD and
Current Reader chains, including long-term stability tests with
reference elements and calibration measurements over the operating
range. A complete metrological characterization of all industrial
input configurations remains part of future work.
}

\arxivnumber{arXiv:2603.14913}

\begin{document}
\maketitle
\flushbottom

\section{Introduction}
Many nuclear-physics experiments require, in addition to the physics readout chain, reliable control of auxiliary parameters that affect apparatus stability and data quality.
Typical quantities include temperature and humidity close to detectors, pressures and vacuum levels along beam lines, as well as currents and diagnostic signals needed for facility operation.
In this context, a slow-control system must be robust, reconfigurable, and easily integrated into laboratory infrastructures, while maintaining consistent metrological performance across heterogeneous channels and ensuring time traceability of operating conditions.
The system presented here was initially developed for the 8Be--X17 experiment at the Legnaro National Laboratories and was subsequently generalized as a reusable platform for experiments with similar requirements.

A central design motivation is to avoid ad-hoc solutions that are poorly documented and difficult to maintain: the goal is a compact, reproducible, modular unit where acquisition hardware, bus management, and telemetry to a time-series backend are integrated coherently.
Adopting a time-series database and web dashboards also enables continuous historical records of slow parameters, simplifying diagnostics, correlations, and the identification of anomalous conditions during operation shifts \cite{GraphiteOverview,GrafanaIntro}.

\section{System overview}
\begin{figure}[t]
  \centering
  \includegraphics[width=0.92\textwidth]{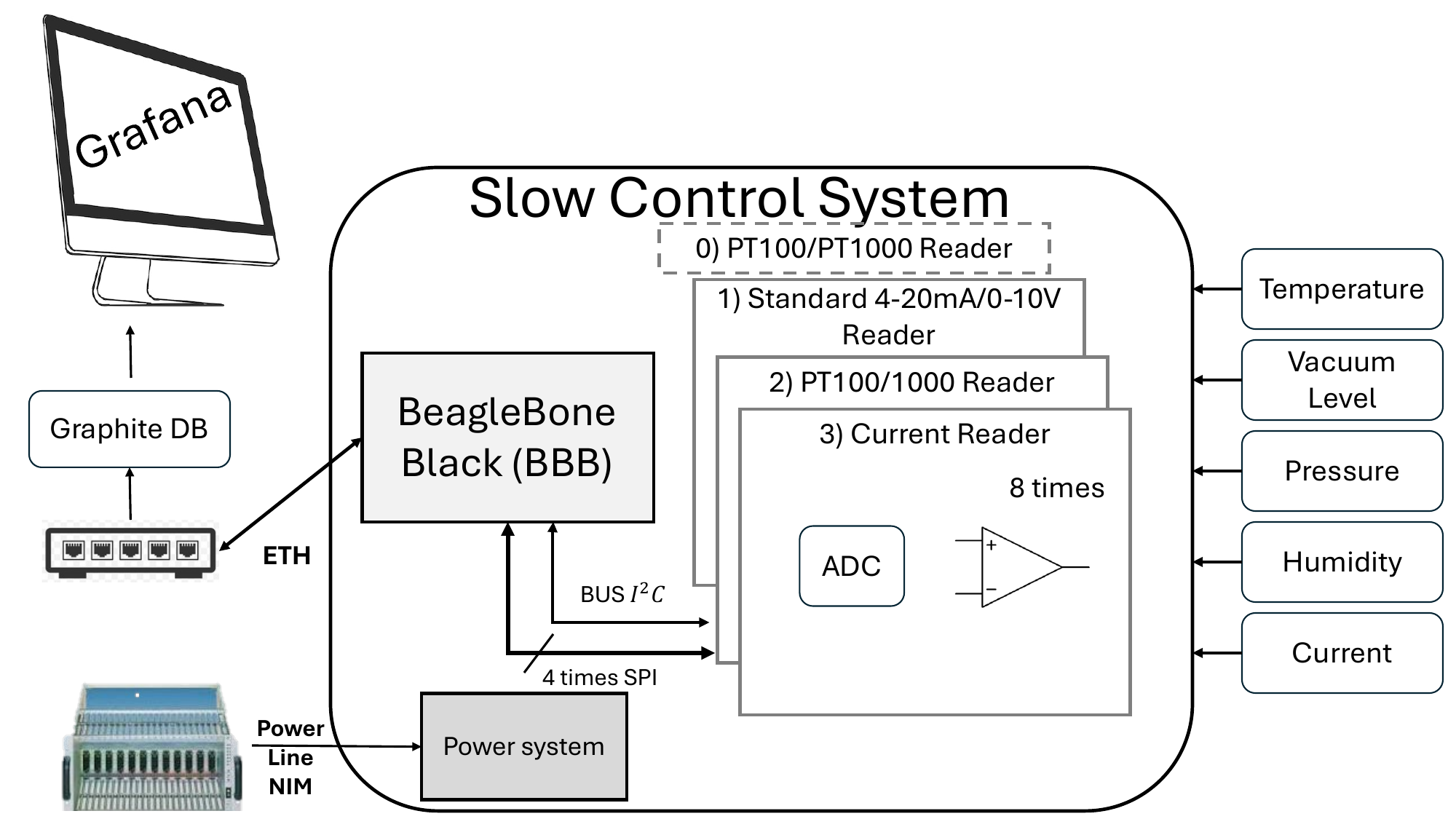}
  \caption{Block diagram of the slow-control module: base board (slot 0), up to three plug-in extension boards (slots 1--3), embedded controller, Ethernet telemetry, and time-series visualization backend.}
  \label{fig:system_overview}
\end{figure}

As shown in figure~\ref{fig:system_overview}, the complete slow-control system is implemented in a standard NIM 2U module that integrates a base controller board and up to three internally installed extension boards.
The base board provides eight native channels for reading PT100/PT1000 RTD probes and hosts the conversion architecture and bus distribution toward extensions.
All boards share the same acquisition philosophy: 16-bit SAR conversion with an external \SI{2.5}{V} reference, so that the metrological meaning of a digital sample is uniform regardless of channel type \cite{AD7689,REF192}.
The AD7689 is a multichannel SAR ADC with SPI interface and an input range of 0--V$_{\mathrm{REF}}$ in unipolar mode, compatible with an analog chain referred to ground and with full-scale set by the reference \cite{AD7689}.
The AD7689 was selected based on previous experience gained in several
instrumentation projects, where it demonstrated reliable operation,
good long-term stability, and performance fully compatible with the
requirements of slow-control applications. Its eight-channel
multiplexed architecture, simple SPI interface, and continued
commercial availability make it a practical and robust choice for this
type of system. While newer SAR ADC families are available, the performance of the
present system is not constrained by the converter itself, and the
modular acquisition architecture would allow future migration to
alternative devices with minimal impact on the overall system design.

Acquisition and networking are handled by a COTS embedded controller (BeagleBone Black), chosen for its extensive documentation and for the possibility of pin-to-pin replacement within the same module family should higher compute performance or a platform update be required \cite{BeagleBoneBlackSRM}.

The BeagleBone Black also integrates two Programmable Real-time Units
(PRUs), which can be used for deterministic low-latency tasks.
In the present implementation, the acquisition process is entirely
handled by the ARM processor through standard Linux services. The sampling rates and timing requirements of the
slow-control application do not impose strict real-time constraints,
and therefore the use of the PRUs was not required.
Nevertheless, the availability of the PRUs represents a valuable
resource for future developments. Potential applications include the
management of deterministic trigger signals, low-latency event
handling, or the implementation of synchronization mechanisms with
strict timing requirements, should future experimental deployments
require such functionality.

External communication is via Ethernet; data are published as time series to Graphite and made available through Grafana dashboards, using a telemetry chain based on well-established open-source tools.
Graphite documentation describes the Carbon/Whisper/graphite-web architecture and the publication of numeric metrics, while Grafana provides a web-based visualization and querying layer suitable for monitoring and operational analysis \cite{GraphiteOverview,GrafanaIntro}.

\section{Mechanical structure and connection interfaces}
\begin{figure}[t]
  \centering
  \begin{minipage}[t]{0.32\textwidth}
    \centering
    \includegraphics[width=\linewidth]{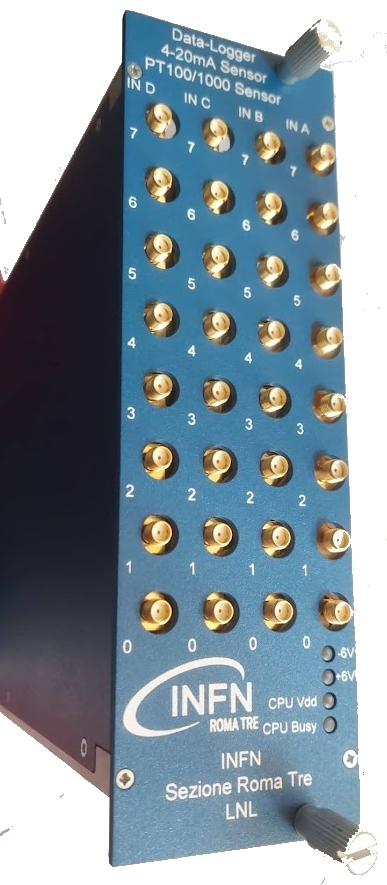}
  \end{minipage}\hfill
  \begin{minipage}[t]{0.64\textwidth}
    \centering
    \includegraphics[width=\linewidth]{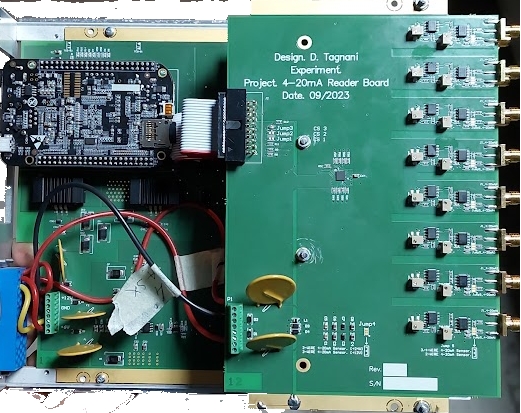}
  \end{minipage}

  \caption{(Left) Front-panel view with SMA connections. (Right) Internal view of the module with the base board and extension slots.}
  \label{fig:module_views}
\end{figure}

The system is built in standard NIM 2U format with a front panel dedicated to sensor connections.
All available channels, both on the base board and on extension boards, are brought to front-panel SMA connectors, providing mechanical robustness and shielded connections.
The SMA interface standardizes terminations and simplifies the construction of repeatable, easily replaceable cabling, which is important during commissioning and maintenance.
The front panel also provides immediate verification of the main supply rails and BBB activity via four dedicated green LEDs: +6V, --6V, CPU\_Vdd (3.3V), and CPU\_Busy, as shown in figure~\ref{fig:module_views}.
These indicators are intended only for immediate visual verification
during installation and troubleshooting. Operational monitoring of the
module health is instead performed through dedicated telemetry channels,
including the internal regulated supply rails and board temperature,
which are continuously acquired and archived by the slow-control system.

Front-panel labeling ensures unambiguous channel identification and correspondence with software numbering.
Columns represent boards, from INA (base controller board, SPI0) to IND (extension board \#3, SPI3), while numbers identify channels within each board.

Internally, a backplane and flat-cable wiring distribute supplies and digital buses, and the slot layout allows rapid replacement of extension boards depending on experimental needs.

\section{Base controller board electronics}
\begin{figure}[t]
  \centering
  \includegraphics[page=1,width=0.95\textwidth]{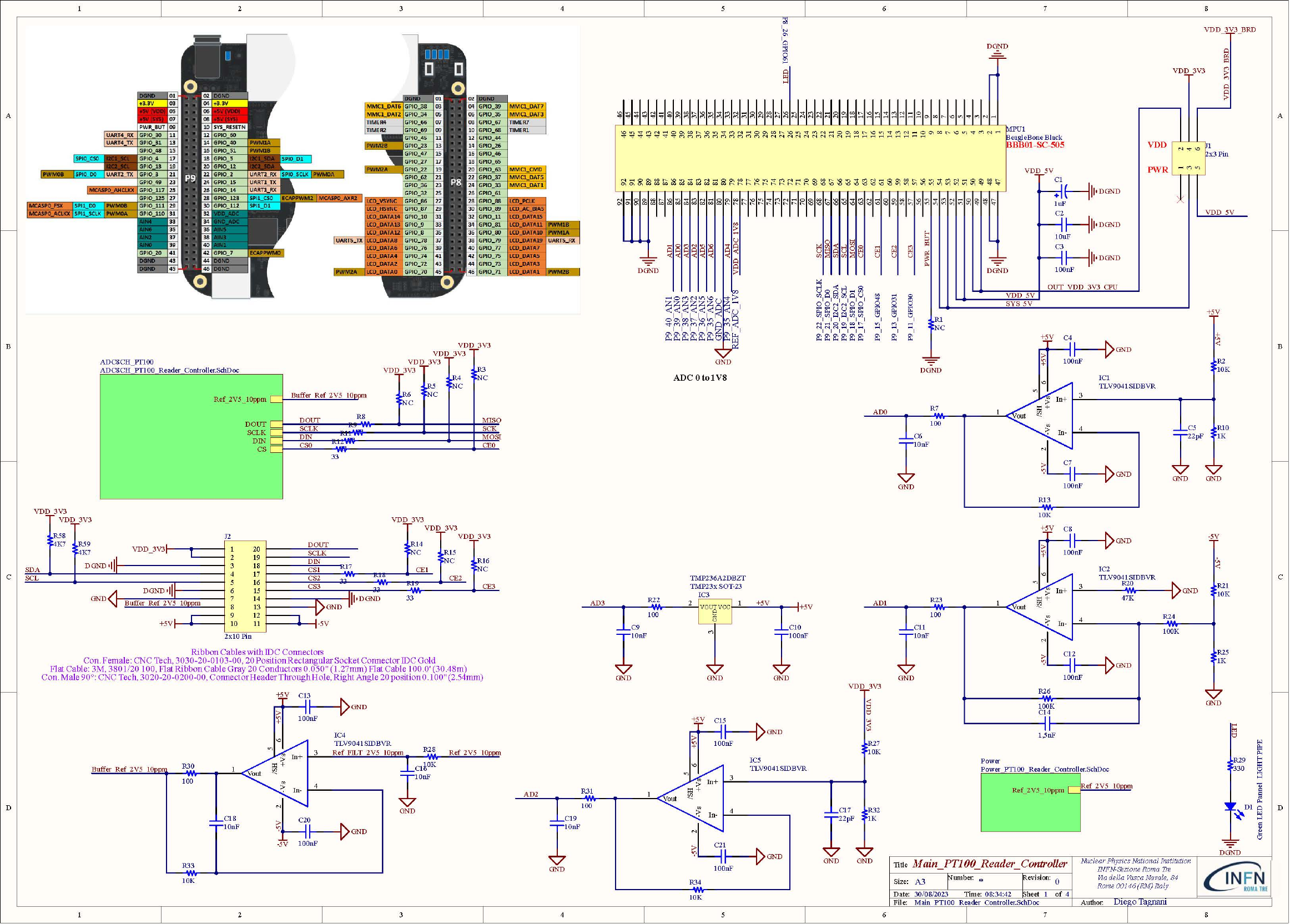}
  \caption{Base board: main schematic (controller and digital interfaces).}
  \label{fig:schem_ctrl_main}
\end{figure}

The base board is the central node of the system and integrates, on a single PCB, three functions that were deliberately combined to reduce cabling complexity and single points of failure: supply conditioning from the NIM crate, the digital interface to the COTS embedded controller (BeagleBone Black), and the measurement chain for eight PT100/PT1000 RTD probes.
The complete schematic of the board and its interconnections to the BBB is shown in figure~\ref{fig:schem_ctrl_main}. \cite{BeagleBoneBlackSRM}

The power subsystem receives the standard crate supplies and conditions them to generate the rails required by the base board and, via the internal connector, by the extension boards.
It includes input protection (resettable PTC fuses and protection diodes) and dedicated linear regulators for local service rails.
In particular, the board generates a regulated \SI{5}{V} rail using a high-current linear regulator and a \SI{3.3}{V} logic rail via a dedicated regulator; these rails power both the base-board logic and the internal backplane distribution toward the extensions. \cite{REF192,AD7689}
The availability of symmetric analog supplies (e.g.\ \SI{\pm5}{V}) also allows proper powering of analog conditioning stages for RTD sensors, preserving dynamic headroom and mitigating limitations associated with single-supply operation when amplification and offset stages are used.

Analog and digital grounds are kept separated on the PCB to reduce ground-loop coupling and to minimize the injection of digital return currents into sensitive low-noise analog front-ends.
The analog (GND) and digital (DGND) planes are routed on distinct copper regions and are tied together only in the power-conditioning area through a low-impedance star point, so that switching currents from bus activity do not flow through the analog front-end region.
This mixed-signal partitioning is consistent with the ADC design intent, where analog conversion performance benefits from a controlled separation of analog reference/return and digital interface return paths \cite{AD7689}.
In this system the analog ground also provides a low-impedance reference for the buffered \SI{2.5}{V} reference distribution, while DGND is used to return and shield the fast digital bus lines routed toward the BBB.

The core of the analog-to-digital conversion is a 16-bit SAR ADC (AD7689), also used on the extension boards to maintain conversion uniformity; with an external \SI{2.5}{V} reference the bit weight is \SI{38.15}{\micro V} \cite{AD7689}, as shown in figure~\ref{fig:schem_ctrl_adc}.
The voltage reference is implemented with a REF192 at \SI{2.5}{V}; its documentation specifies a maximum temperature coefficient of 5~ppm/\si{\celsius} and line/load regulation figures that help stabilize the ADC scale over time \cite{REF192}.
In the circuit, the REF192 reference is filtered and buffered before
distribution to the ADC section and, through the internal bus, to all
extension boards. A single REF192 device located on the base controller
board therefore provides the master \SI{2.5}{V} reference for the entire
system. The reference is distributed through the internal backplane
connector and is used directly by the AD7689 converters on the extension boards; no local voltage references are generated on the extension boards. This
architecture guarantees a common and reproducible metrological scale
across all installed modules and minimizes potential channel-to-channel
gain differences arising from multiple independent references.

The RTD measurement chain is designed to convert the PT100/PT1000 resistance variation into a signal compatible with the ADC unipolar input range (0--V$_{\mathrm{REF}}$).
Specifically, the measurement uses a high-stability balanced Wheatstone bridge powered by the precision reference and an offset-cancellation network that enables single-supply readout while keeping the operating point within 0--\SI{2.5}{V}.
This approach enables use of the AD7689 in unipolar mode and maximizes usable dynamic range over the full operating span \SI{-200}{\celsius} $\rightarrow$ \SI{+300}{\celsius}, avoiding bipolar conversion and reducing potential drift sources associated with virtual references or software full-scale shifts. \cite{AD7689}

From the digital-architecture perspective, the BBB is the control node of the system: it handles converter readout, channel configuration, and data publication.
The BBB is directly connected to the base board via SPI lines and service lines (I\textsuperscript{2}C and GPIO) using dedicated pins on connector P9; the main SPI lines (SCLK, MOSI, MISO, and chip-select) are routed to the base-board ADC and, through the internal bus, to the extension boards, with physically separated selection lines for slots that allow addressing installed modules and reducing coupling between parallel acquisitions.
In this way, the base board acts as a functional backplane: it integrates the power and bus infrastructure, while extension boards primarily implement analog conditioning and A/D conversion for their respective channels, keeping the acquisition and telemetry logic on the embedded controller unchanged.

The connection between the base controller board and the extension boards is implemented through a dedicated interface connector and an internal flat cable.
This interface distributes supplies, the reference voltage, and the digital communication buses required for system operation.
The complete pin sequence and associated function are listed in table~\ref{tab:J2_pinout}.
The base controller board is permanently associated with \texttt{CS0} (slot~0) and therefore does not require any addressing jumpers, while each plug-in extension board carries three on-board SMD solder jumpers that implement a one-hot hardware selection among \texttt{CS1}, \texttt{CS2}, and \texttt{CS3}.
During assembly, exactly one jumper is closed (solder-bridged) to bind the extension board to the corresponding chip-select line, establishing a deterministic hardware-to-software association between physical slot and acquisition addressing.

\begin{table}[t]
\centering
\caption{Pin sequence of the internal connector J2 (2$\times$10) interfacing the extension boards. Net names follow the labels used in the electrical schematics.}
\label{tab:J2_pinout}
\begin{tabular}{@{}r l p{0.62\textwidth}@{}}
\toprule
\textbf{PIN} & \textbf{NET} & \textbf{Description} \\
\midrule
1  & VDD\_3V3               & \SI{3.3}{V} logic supply for digital interfaces and on-board logic on extensions. \\
2  & -5V                    & \SI{-5}{V} analog supply for stages requiring dual supply. \\
3  & +5V                    & \SI{+5}{V} analog supply for conditioning stages. \\
4  & Buffer\_Ref\_2V5\_10ppm & Buffered \SI{2.5}{V} analog reference (common reference distribution among boards). \\
5  & GND                    & Analog ground. \\
6  & GND                    & Analog ground. \\
7  & DGND                   & Digital ground. \\
8  & DGND                   & Digital ground. \\
9  & SCL                    & I\textsuperscript{2}C clock line (auxiliary bus on connector). \\
10 & SDA                    & I\textsuperscript{2}C data line (auxiliary bus on connector). \\
11 & DGND                   & Digital ground. \\
12 & MISO                   & SPI MISO: data from module to embedded controller. \\
13 & SCK                    & SPI clock shared on the slot. \\
14 & MOSI                   & SPI MOSI: data from embedded controller to module. \\
15 & CS1                    & Chip-select/slot-select line (address selection via SMD solder jumper). \\
16 & CS2                    & Chip-select/slot-select line (address selection via SMD solder jumper). \\
17 & CS3                    & Chip-select/slot-select line (address selection via SMD solder jumper). \\
18 & CEn                    & SPI chip enable (enables the selected device on the slot). \\
19 & DGND                   & Digital ground to reduce loops and coupling on fast signals. \\
20 & DGND                   & Digital ground to reduce loops and coupling on fast signals. \\
\bottomrule
\end{tabular}
\end{table}

\begin{figure}[t]
  \centering
  \includegraphics[page=2,width=0.95\textwidth]{figures/Schematics_Controller.PDF}
  \caption{Base board: 8-channel 16-bit ADC section (AD7689) and reference distribution.}
  \label{fig:schem_ctrl_adc}
\end{figure}

\section{Extension boards}
The extension boards implement specialized functional modules and up to three units can be installed in addition to the base board.
Each board provides eight channels, uses a 16-bit AD7689 ADC with a \SI{2.5}{V} reference, and interfaces to the embedded controller via one of the three SPI lines dedicated to slots 1--3, preserving a coherent conversion chain across all signal types \cite{AD7689,REF192}.
The I\textsuperscript{2}C bus is available on the internal connector for auxiliary functions and is currently used, at the functional level, only in the Current Reader module for gain-range selection.

The PT100/PT1000 extension board increases the number of RTD channels while preserving the same philosophy as the base board.
Each channel uses a high-precision, high-stability balanced Wheatstone bridge, with sensor-type selection via jumper (PT100 or PT1000) and offset cancellation to ensure compatibility with single-supply readout and a unipolar 0--\SI{2.5}{V} ADC input range.
Using a balanced bridge for resistive sensors is a well-established technique: the sensor resistance variation unbalances the bridge and generates a differential signal, while the common-mode level can be managed to avoid saturating the measurement stage \cite{ADInAmpGuide}.
Within the system, the choice of reference and ADC maintains a uniform metrological scale between base and extension channels, while trimming networks allow offset and gain adjustment over the \SI{-200}{\celsius} $\rightarrow$ \SI{+300}{\celsius} range with a design accuracy at the \SI{1}{\%} level \cite{REF192,AD7689}.

\begin{figure}[t]
  \centering
  \includegraphics[page=3,width=0.95\textwidth]{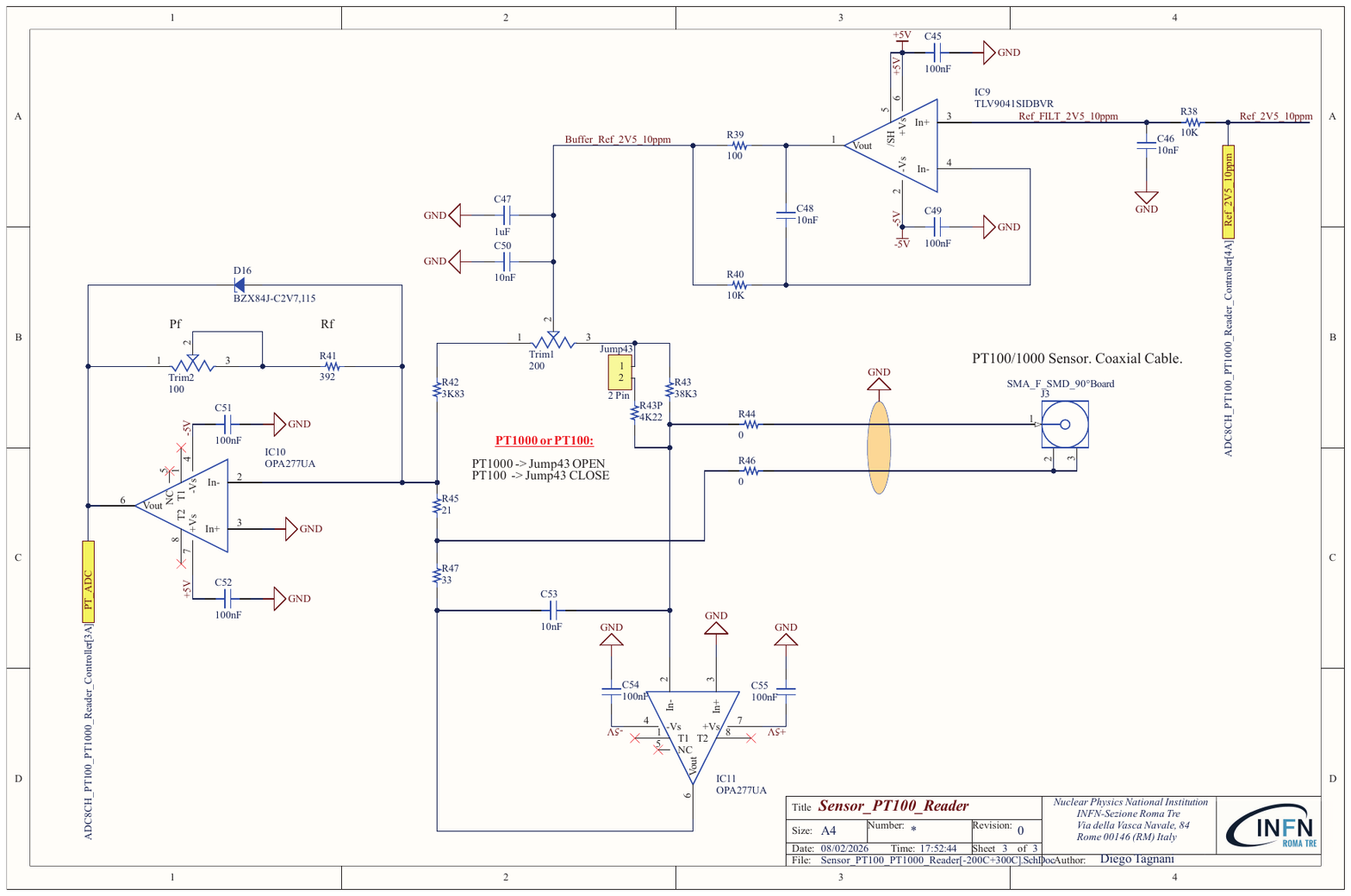}
  \caption{Schematic of the PT100/PT1000 extension board (eight channels, PT100/PT1000 selection via jumper).}
  \label{fig:schem_ctrl_ext_pt100_pt1000}
\end{figure}

The schematic of the PT100/PT1000 extension board is shown in figure~\ref{fig:schem_ctrl_ext_pt100_pt1000}.

The 4--20~mA / 0--10~V extension board is dedicated to standard industrial signals, typically provided by process transducers (pressure, level, flow) and environmental sensors.
The 4--20~mA current loop is widely adopted for its robustness to noise and the ``live zero'' principle, which allows distinguishing a near-zero value from a disconnection or line fault.
Industrial documentation also describes several wiring options for 4--20~mA transmitters, including 2-wire (loop-powered), 3-wire, and 4-wire configurations, which differ in how power and signal lines are separated.
In the presented system, the board supports both 2-wire and 3/4-wire sensors and provides a selectable \SI{12}{V} or \SI{24}{V} sensor supply, enabling compatibility with transducers requiring different supply levels without changing the basic cabling.
For loop-powered 2-wire transmitters, the module can directly provide
the selected \SI{12}{V} or \SI{24}{V} supply voltage through the same
current loop used for signal transmission. The board is also
compatible with conventional 3-wire and 4-wire transmitters, for which
sensor power and signal connections remain separate. This arrangement
allows direct interfacing with the most common industrial
instrumentation standards without requiring additional interface or
signal-conditioning hardware.

For the 0--10~V channel, the interface is simple and widely compatible but generally more sensitive to interference and voltage drops on long runs than current transmission; including both standards in a single module therefore increases integration flexibility.
The desired standard is selected manually per channel via board jumpers.
The conditioned signal is digitized by the on-board 16-bit ADC, ensuring conversion consistency with the other system modules \cite{AD7689}.

\begin{figure}[t]
  \centering
  \includegraphics[page=3,width=0.95\textwidth]{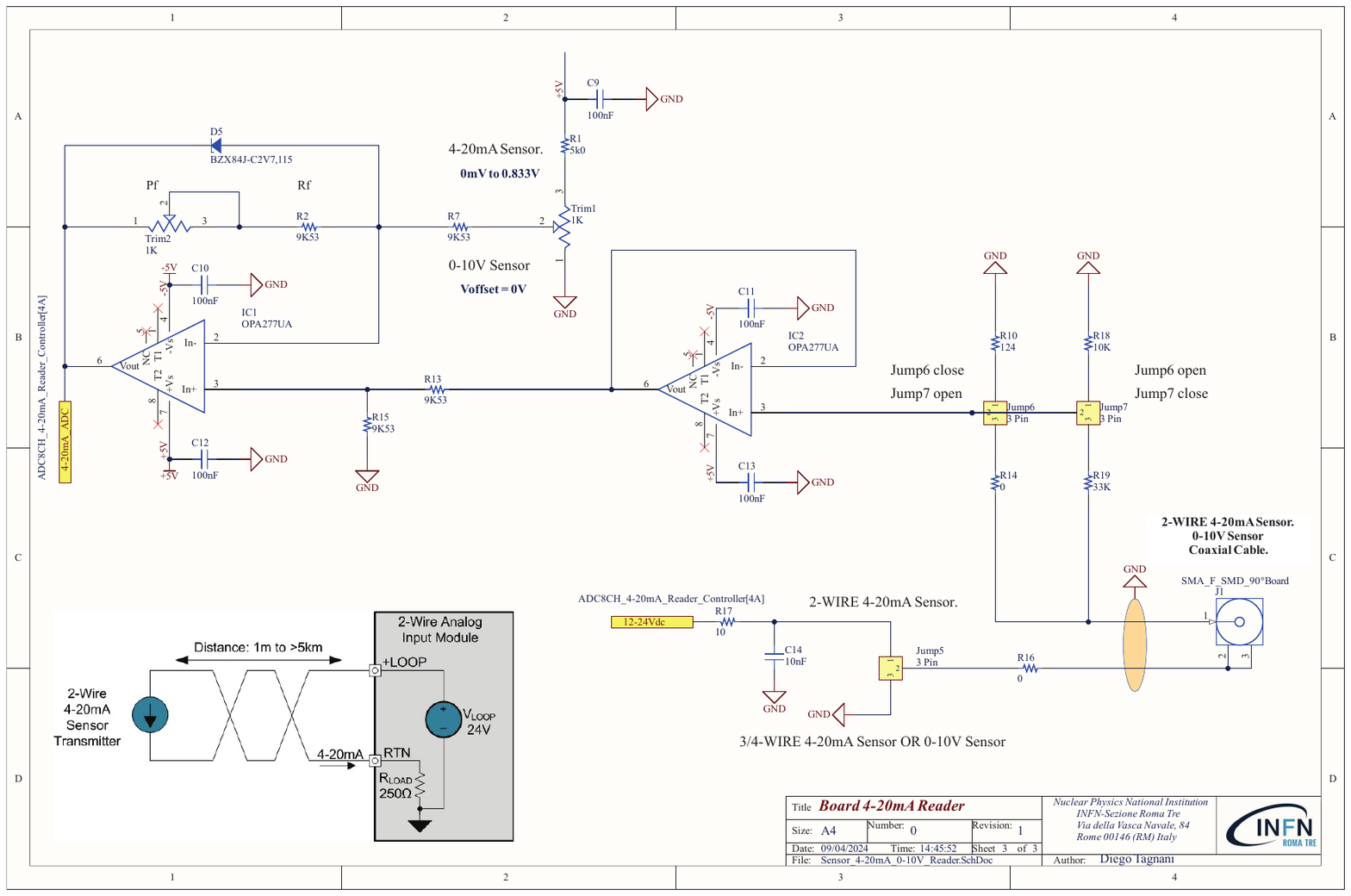}
  \caption{Schematic of the 4--20~mA / 0--10~V extension board (eight channels, manual per-channel selection via jumper).}
  \label{fig:schem_ext_4_20mA_0_10V}
\end{figure}

The schematic of the 4--20~mA / 0--10~V extension board is shown in
figure~\ref{fig:schem_ext_4_20mA_0_10V}.

The Current Reader board is dedicated to current measurements in the \SI{10}{nA}--\SI{150}{\micro A} range, useful for diagnostics and alignment on beam lines or for monitoring transported beam-line current.
Values below the nominal lower bound may be measurable, but require dedicated experimental validation before being considered an operational specification.
The analog chain implements a current-to-voltage conversion with selectable gain range, and automatic gain-range selection is managed locally via the I\textsuperscript{2}C bus.
The operational motivation is to maintain useful resolution over multiple decades: without autoranging, a single gain setting would either saturate at high values or provide poor sensitivity at low values.
If the expected current range is well known and stable over time, manual range selection is also possible; this allows maximizing the measurable range for the selected gain at the cost of potential saturation or reduced sensitivity if conditions change.
The resulting signal is digitized by a 16-bit ADC, preserving acquisition uniformity and integration into the common backend \cite{AD7689}.
Overall, the board provides a measurement channel usable both during setup/alignment and for continuous monitoring, while keeping the same mechanical (front-panel SMA) and digital (controller bus) interfaces as the rest of the system.

\begin{figure}[t]
  \centering
  \includegraphics[page=3,width=0.95\textwidth]{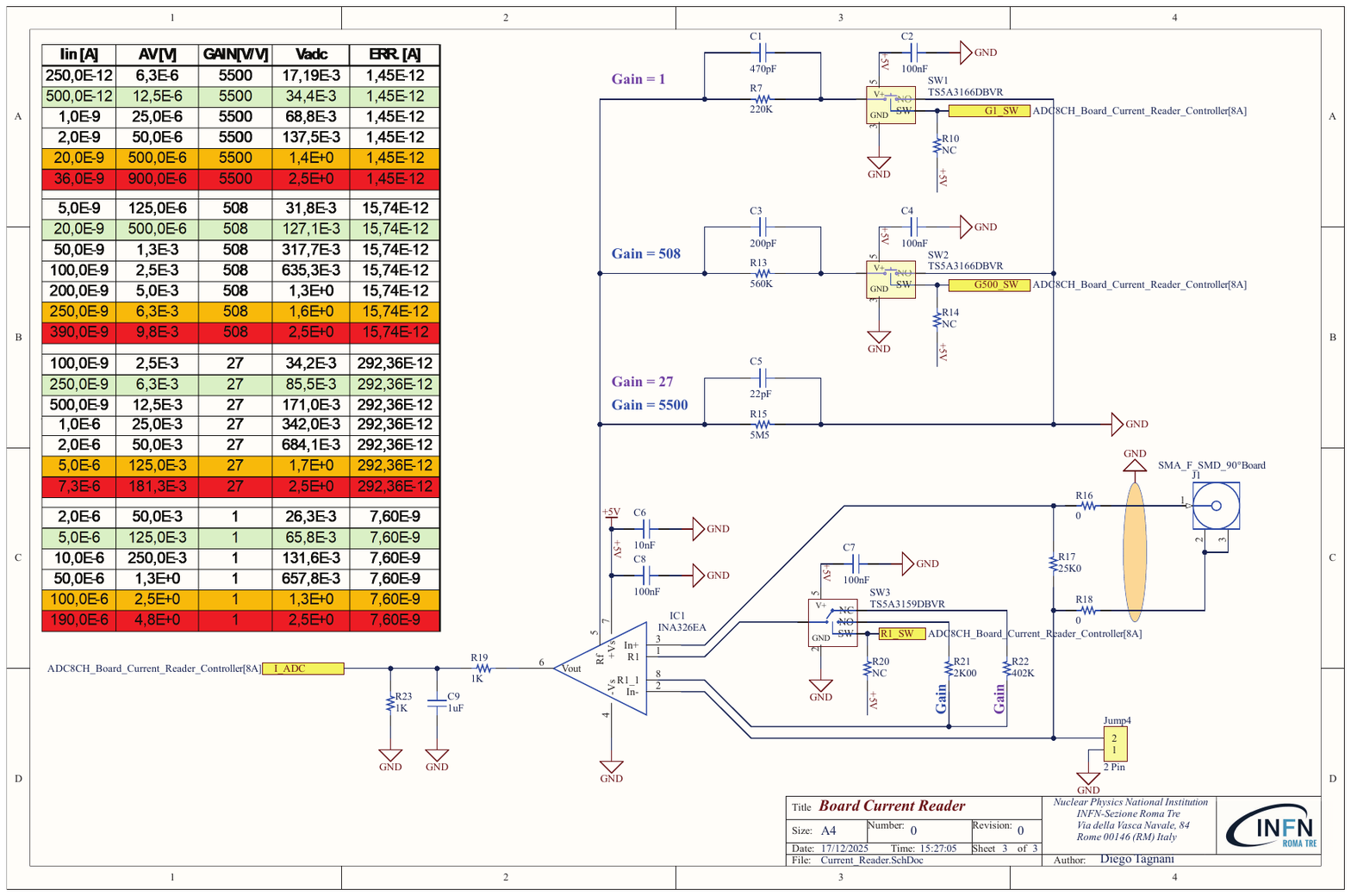}
  \caption{Schematic of the Current Reader board (eight channels, automatic gain selection and 16-bit digitization).}
  \label{fig:schem_ctrl_ext_current_reader}
\end{figure}

The schematic of the Current Reader extension board is shown in
figure~\ref{fig:schem_ctrl_ext_current_reader}.

\section{Data Acquisition, Publication, and Visualization}
The software chain was designed to enable reproducible slow-control operation in an experimental environment, transforming a heterogeneous set of channels (RTD temperatures, industrial signals, and diagnostics) into time-series telemetry that can be queried both in real time and offline.
A guiding principle is the separation of responsibilities among acquisition, control, and presentation, in order to reduce coupling between components and simplify long-term maintenance and evolution.
Accordingly, the system is organized into three layers: (i) an acquisition firmware that configures the hardware, runs the measurement cycle, and generates metrics, (ii) an HTTP control layer that acts as a bridge between the network and the device, and (iii) a web user interface for configuration and supervision \cite{GraphiteOverview,GrafanaIntro}.
This architecture allows evolving the UI and run-control mechanisms without altering the acquisition logic and without introducing dependencies on the presentation environment.

Measurements are published to the backend through Graphite, using the Carbon service as the ingestion and buffering point for time series.
Carbon's \emph{plaintext} protocol encodes each sample as \texttt{<metric path> <metric value> <metric timestamp>}, providing a minimal and robust interface for sending numeric metrics over the network \cite{GraphiteFeedingCarbon}.
Time traceability is ensured by assigning the timestamp at the datalogger level on the BBB.
The BBB synchronizes its system clock over the network using standard NTP/SNTP services when Ethernet connectivity is available \cite{RFC5905,TimesyncdMan}.
Consequently, every stored data point carries its acquisition time, and the same timestamp is used by Grafana to render time axes and correlate channels and modules consistently \cite{GraphiteFeedingCarbon,GrafanaIntro}.
On the backend, Carbon manages reception and caching of recent data and enables persistence to the configured storage, making time series available for queries and rendering \cite{GraphiteOverview}.
Grafana is used downstream as a web front-end for operational monitoring, historical navigation, and alerting integration, enabling dashboards consistent with control-room workflows and diagnostic needs during data taking \cite{GrafanaIntro}.
The adoption of established time-series tools also helps standardize telemetry across different subsystems, reducing integration overhead when combining heterogeneous sensors or when modules are installed in different configurations.

A key aspect in an experimental environment is the definition of a
stable and semantically meaningful channel-identification scheme.
To achieve this objective, the complete system configuration is
performed through a web interface served directly by the datalogger,
without requiring direct access to the operating system or manual
modification of configuration files.

\begin{figure}[t]
\centering
\includegraphics[width=0.95\textwidth]
{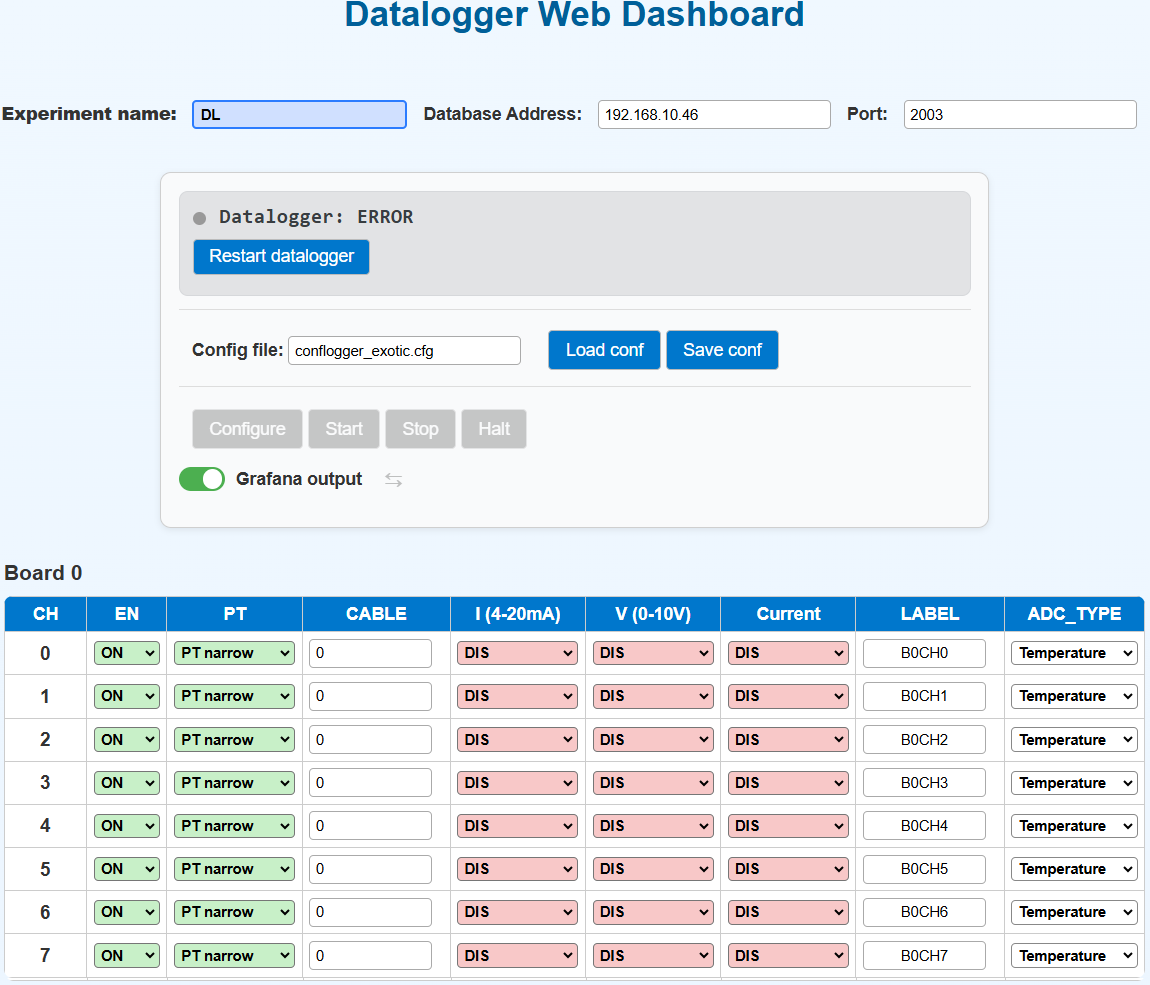}
\caption{
Main configuration page of the datalogger web interface.
The user can define the experiment name, Graphite server address and port,
enable or disable telemetry publication, and manage configuration files.
}
\label{fig:gui_main}
\end{figure}

The web interface allows the user to define the experiment identifier,
the destination Graphite server address and port, and whether telemetry
publication is enabled. 
The main configuration page is shown in figure~\ref{fig:gui_main}.
In addition, every acquisition channel of the
installed boards can be individually configured according to the
selected hardware module and experimental requirements.

\begin{figure}[t]
\centering
\includegraphics[width=0.90\textwidth]
{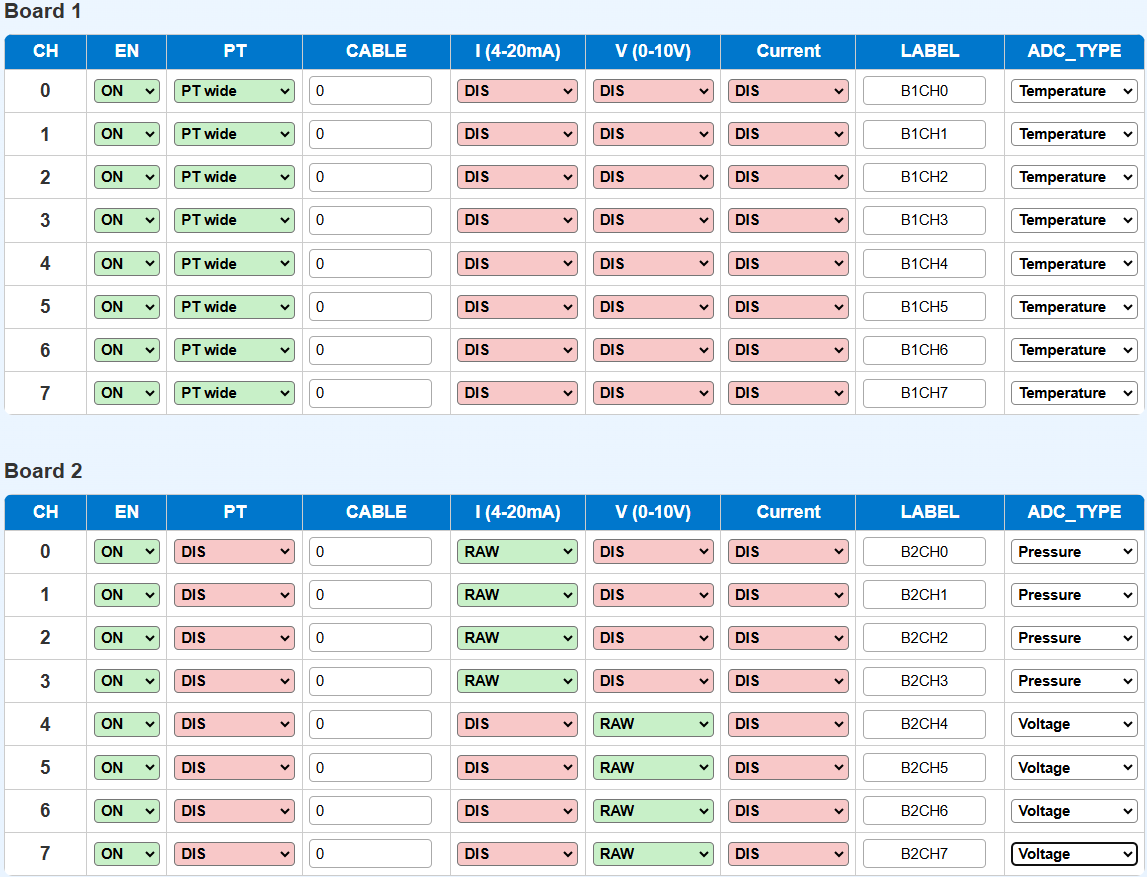}
\caption{
Channel configuration page for extension boards. For each channel the
user can enable or disable acquisition, select the measurement mode,
assign a semantic channel label (\texttt{LABEL}), and define the
corresponding measurement class (\texttt{ADC\_TYPE}). The example
shows channels configured for temperature, pressure, and voltage
measurements.
}
\label{fig:gui_channels_ext}
\end{figure}

\begin{figure}[t]
\centering
\includegraphics[width=0.90\textwidth]
{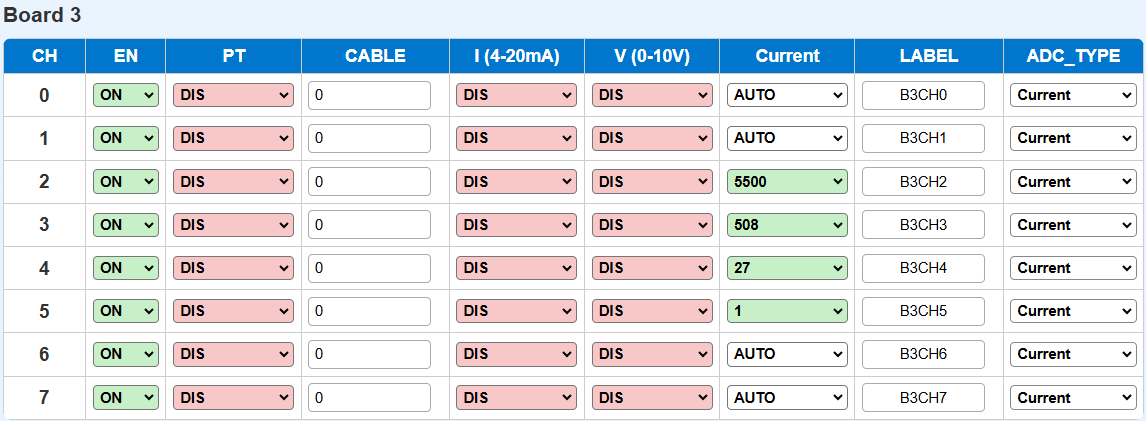}
\caption{
Channel configuration page for the Current Reader extension board.
The interface allows selection of the current-measurement range,
channel activation, logical channel labels (\texttt{LABEL}), and
measurement-class identifiers (\texttt{ADC\_TYPE}).
}
\label{fig:gui_channels_current}
\end{figure}

For each channel, the user can enable or disable data acquisition,
select the measurement mode, assign a descriptive channel label
(\texttt{LABEL}), and define the corresponding measurement class
(\texttt{ADC\_TYPE}), such as temperature, pressure, voltage, or
current. These parameters determine both the acquisition behavior and
the hierarchical metric name published to the telemetry backend.
Channels marked as disabled are excluded from
telemetry publication and are therefore not stored in the database.
Configuration files can be saved locally on the datalogger and
reloaded when needed, allowing complete experiment-specific setups to
be preserved and rapidly restored across different experimental
campaigns.
Representative channel-configuration pages are shown in figures~\ref{fig:gui_channels_ext} and
\ref{fig:gui_channels_current}.

Telemetry is published to Graphite using a hierarchical metric
structure of the form

\begin{center}
\texttt{DLOG.ExperimentName.ADC\_TYPE.LABEL}
\end{center}

where \texttt{DLOG} is a fixed system prefix, while
\texttt{ExperimentName}, \texttt{ADC\_TYPE}, and
\texttt{LABEL} are defined by the user through the web interface.

This naming convention decouples the logical identification of a
measurement channel from the physical board, slot, and channel
numbering, preserving semantic information independently of the
hardware layout and allowing the same hardware to be reconfigured for
different experiments without requiring changes to dashboard
definitions or analysis tools.
As a result, the same module can be reused in different experiments
while maintaining human-readable channel names, consistent telemetry
organization, and reusable Grafana dashboards with minimal
reconfiguration effort.

In addition to the metric naming scheme, channel-level configuration
provides flexibility across different experimental deployments: the
user defines which channels are active and how they are interpreted
(input type and specific parameters), without requiring hardware
changes.
The approach favors operational consistency: for each channel, a single physical decoding mode is selected, reducing ambiguity and preventing misconfiguration when different modules or cabling are present.
In parallel, a portion of telemetry can be dedicated to service quantities (internal diagnostics and operating conditions), which are stored as time series and can thus be correlated in time with the quantities of interest, increasing the ability to distinguish real sensor drifts from system-induced variations.

Data retention is managed through the Graphite backend and the Whisper
time-series database. In the current deployment, measurements are
stored with a resolution of 10~s for the first 6~h after acquisition.
Older samples are automatically consolidated to a resolution of
1~min for the following 6~days, and subsequently to a resolution of
10~min for a retention period of approximately 1800~days
(about 5~years).

This multi-resolution storage strategy optimizes disk-space usage while
preserving the information required to analyze both short-term
variations and long-term trends of the monitored quantities. The
selected retention policy is adequate for the typical operating
conditions of the system, where data-taking campaigns generally span
from days to weeks.

At the end of a measurement campaign, data can be exported from the
Graphite/Grafana environment and archived according to the specific
requirements of each experiment. Future modifications of the retention
policy and archival strategy can be implemented through Graphite
configuration files without requiring changes to the datalogger
hardware or acquisition software.

Graphite is designed as an enterprise-ready time-series monitoring stack and is commonly deployed to store and render large metric sets \cite{GraphiteApp,GraphiteOverviewDoc}.
While a single module is limited to slot~0 plus three extension slots by the available dedicated SPI chip-select lines, the overall system scales by operating multiple modules in parallel and pushing all metrics to a common Graphite backend, provided each module uses a unique metric namespace (e.g.\ crate/module identifiers in the metric path) \cite{GraphiteFeedingCarbon,GraphiteOverviewDoc}.

\section{Performance and calibrations}
\subsection{RTD chain (PT100/PT1000): precision and stability}

The temperature chain was verified both under reference conditions and under realistic operating conditions.

In the first case, the RTD inputs were terminated with precision resistors
(nominal \SI{100}{\ohm}, \SI{95.3}{\ohm}, \SI{107.2}{\ohm},
\SI{18.493}{\ohm}, \SI{212.019}{\ohm},
\SI{1000}{\ohm}, \SI{953}{\ohm}, and \SI{1072}{\ohm})
to reproduce known operating points and isolate the electronic
contribution from probe and environmental effects.

For long-term stability measurements, precision metal-film reference
resistors of \SI{100}{\ohm} and \SI{108}{\ohm}
(tolerance $\pm$0.1\%, temperature coefficient
$\pm$5 ppm/\si{\celsius}) were connected to dedicated acquisition
channels. These values correspond approximately to PT100 operating
points at \SI{0}{\celsius} and \SI{20.7}{\celsius}, respectively.

The measurements were continuously acquired over several days and
processed through the complete acquisition chain, including the analog
front-end, ADC conversion, embedded processing, telemetry publication,
and database storage, thereby validating the stability of the entire
measurement system rather than that of the analog electronics alone.

\begin{figure}[t]
\centering
\begin{minipage}{0.49\textwidth}
\centering
\includegraphics[width=\linewidth]
{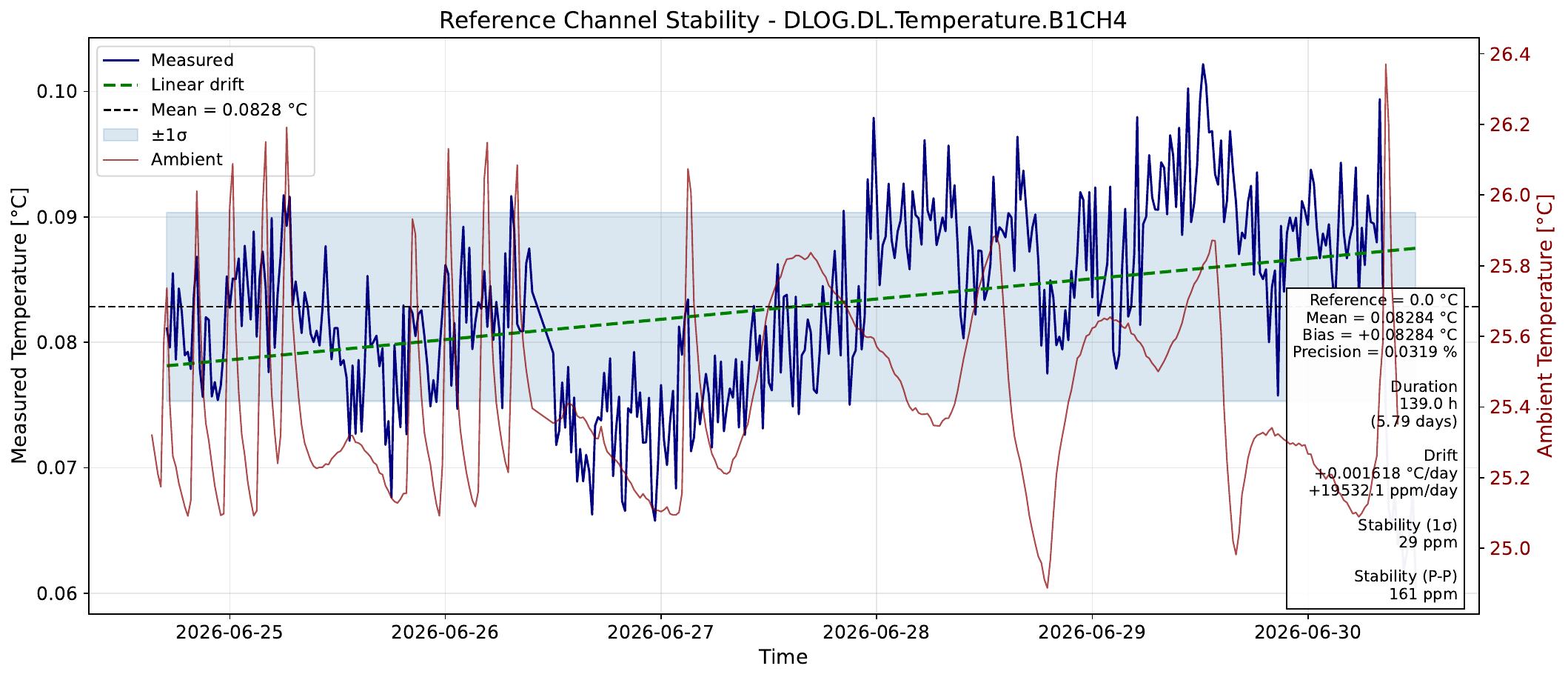}
\end{minipage}
\hfill
\begin{minipage}{0.49\textwidth}
\centering
\includegraphics[width=\linewidth]
{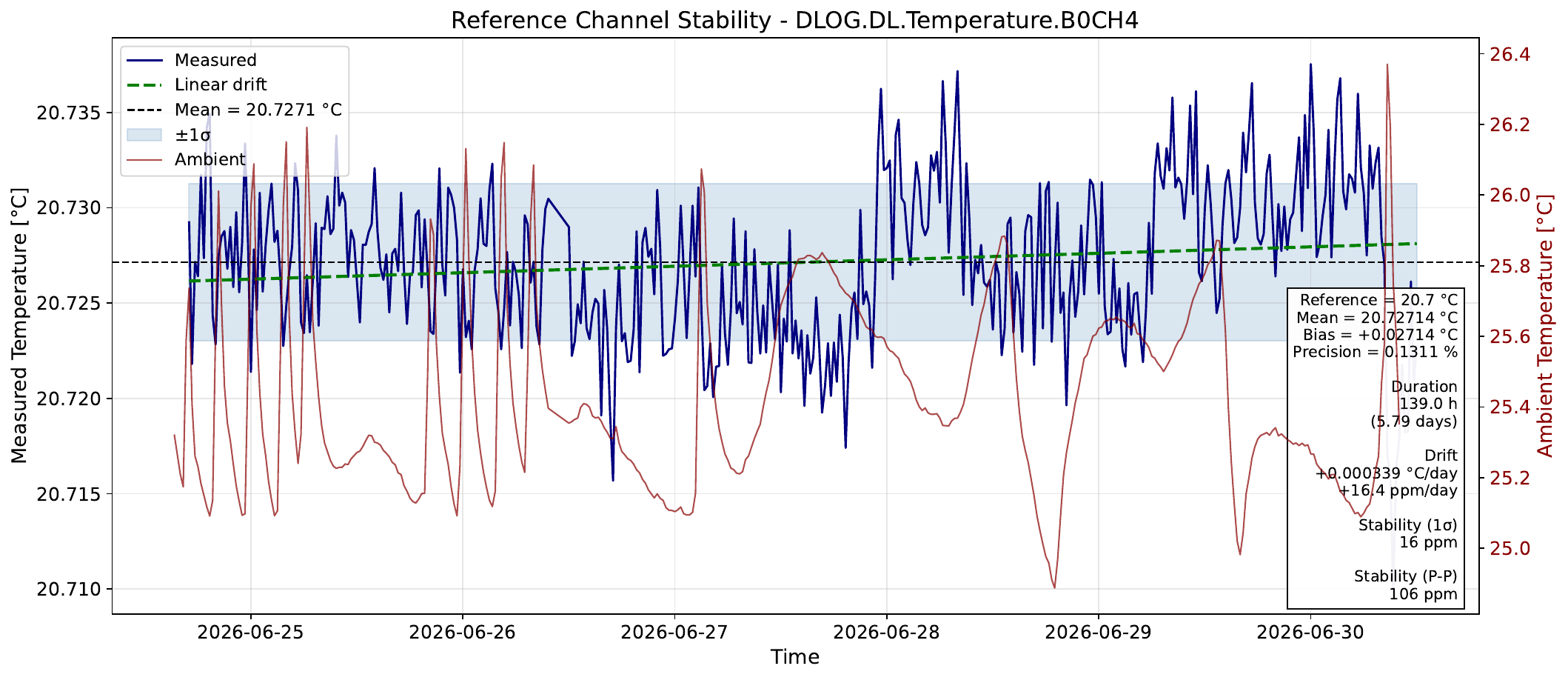}
\end{minipage}
\caption{
Long-term stability measurements of the RTD acquisition chain using
precision metal-film reference resistors ($\pm$0.1\%,
$\pm$5 ppm/\si{\celsius}).
Left: \SI{100}{\ohm} reference resistor
(PT100 equivalent at approximately \SI{0}{\celsius}).
Right: \SI{108}{\ohm} reference resistor
(PT100 equivalent at approximately \SI{20.7}{\celsius}).
}
\label{fig:rtd_stability_reference}
\end{figure}

Figure~\ref{fig:rtd_stability_reference} shows the long-term stability
measurements performed with the \SI{100}{\ohm} and \SI{108}{\ohm}
precision reference resistors. 

For the \SI{100}{\ohm} test, corresponding to a PT100 operating point
of approximately \SI{0}{\celsius}, a mean reading of
\SI{0.08284}{\celsius} was obtained, corresponding to a bias of
\SI{0.08284}{\celsius}. Over an acquisition period of approximately
5.8 days, the measured drift was
\SI{0.001618}{\celsius\per\day},
with a stability of 29~ppm ($1\sigma$) and
161~ppm peak-to-peak.

For the \SI{108}{\ohm} test, corresponding to a PT100 operating point
of approximately \SI{20.7}{\celsius}, a mean reading of
\SI{20.72714}{\celsius} was obtained, corresponding to a bias of
\SI{0.02714}{\celsius}. Over an acquisition period of approximately
5.8 days, the measured drift was
\SI{0.000339}{\celsius\per\day}
(\SI{16.4}{ppm\per\day}), with a stability of
16~ppm ($1\sigma$) and 106~ppm peak-to-peak.

The results obtained at both reference values demonstrate that the
observed stability is not limited to a single operating point and
remains consistent across the temperature range typically encountered
during detector slow-control operation.

These measurements substantially reduce the influence of sensor and
environmental effects, allowing a more direct evaluation of the
intrinsic stability of the analog front-end, precision reference
distribution network, and ADC conversion chain.

Additional measurements performed with PT100 and PT1000 probes under
normal laboratory conditions confirmed correct tracking of ambient
temperature variations and good inter-channel consistency, demonstrating
the suitability of the system for continuous slow-control operation.

\subsection{Current chain: linearity, accuracy, and long-term stability}

The \emph{Current Reader} board was characterized using a dedicated
bench setup designed to cover the full operating range of the module.
Reference currents were generated from a Keithley 487 precision voltage
source through a set of high-value metal-film resistors
(\SI{5}{G\ohm}, \SI{9}{M\ohm}, \SI{7}{M\ohm},
\SI{5}{M\ohm}, and \SI{3}{M\ohm}).
Multiple resistor values were required in order to span the complete
measurement range while remaining within the maximum output voltage
capability of the source.

The generated currents covered the operating range from \SI{10}{nA} to \SI{150}{\micro A}, allowing
verification of both the analog front-end response and the automatic
gain-range selection system implemented on the board.

\begin{figure}[t]
\centering
\includegraphics[width=0.85\textwidth]
{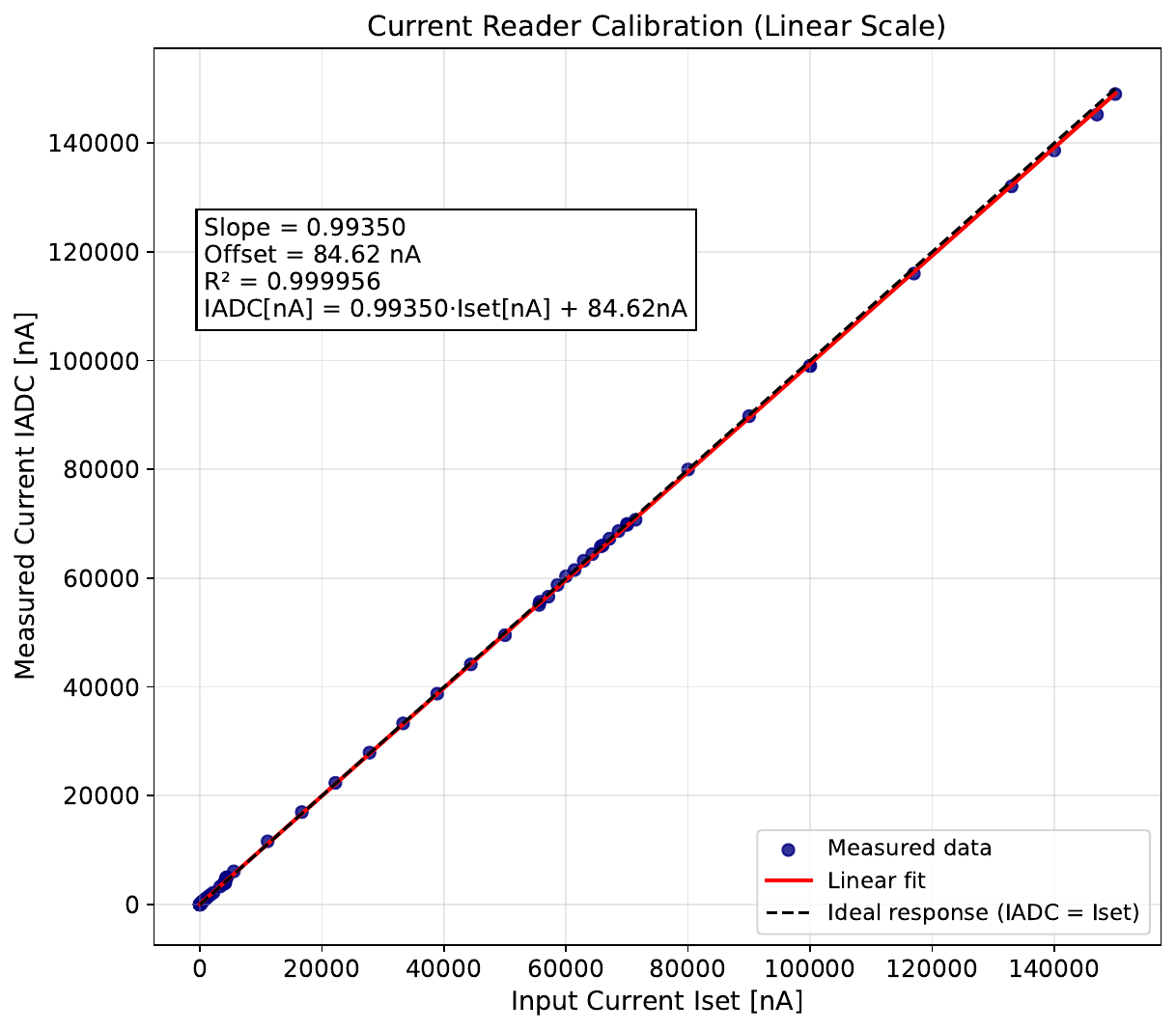}
\caption{
Current Reader calibration over the full measurement range.
Measured data are compared with the ideal response and with the
best-fit linear regression. The characterization includes all
automatic gain-selection regions used by the autoranging system.
}
\label{fig:current_linearity}
\end{figure}

Figure~\ref{fig:current_linearity} shows the measured current as a
function of the applied input current. A linear regression of the
complete dataset yields

\begin{equation}
I_{\mathrm{ADC}} =
0.99350\,I_{\mathrm{Iset}} + 84.62~\mathrm{nA},
\end{equation}

with a coefficient of determination
$R^2 = 0.999956$.

The fitted slope differs from unity by less than \SI{0.65}{\%}, while the equivalent offset is approximately
\SI{85}{nA}. The absence of visible discontinuities in the transfer
curve demonstrates correct operation of the automatic gain-selection
algorithm across the different measurement ranges. The high value of
$R^2$ confirms that the current-measurement chain preserves excellent
linearity across the complete operational range of the instrument.

Long-term stability was evaluated by applying a constant leakage
current generated by the same calibration setup and continuously
acquiring the resulting measurement for approximately
5.7 days.

\begin{figure}[t]
\centering
\includegraphics[width=0.85\textwidth]
{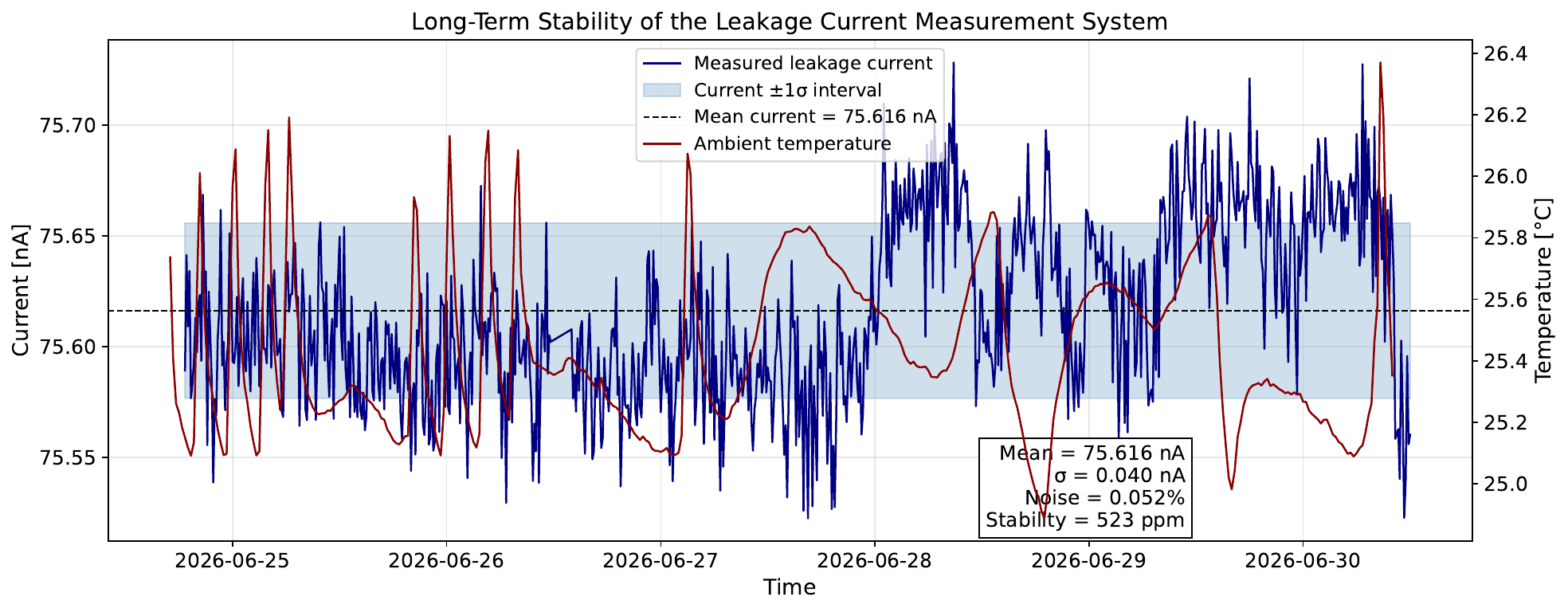}
\caption{
Long-term stability measurement of the Current Reader chain.
The upper trace shows the measured leakage current, while the lower
trace reports the ambient laboratory temperature recorded during the
same time interval. The measurement was acquired continuously for
approximately 5.7 days.
}
\label{fig:current_stability}
\end{figure}

Figure~\ref{fig:current_stability} reports the measured current
together with the ambient laboratory temperature recorded during the
test interval. A mean current of \SI{75.616}{nA} was observed with a
standard deviation of \SI{0.040}{nA}, corresponding to a relative
noise level of approximately \SI{0.052}{\%}. The measured long-term
stability was 523~ppm over the entire acquisition period.

Ambient laboratory temperature varied between
\SI{25.0}{\celsius} and \SI{26.3}{\celsius} during the measurement.
No active temperature stabilization was applied to the test setup,
although the laboratory environment was air-conditioned. Despite these
environmental variations, the measured current remained stable
throughout the acquisition period, demonstrating the suitability of
the system for continuous long-duration monitoring applications.

The present characterization focused on practical operating
conditions representative of laboratory deployments. While the
laboratory environment was air-conditioned, no dedicated humidity
monitoring was performed during the measurements. A systematic study
of temperature and humidity dependencies under controlled
environmental conditions remains part of future work.

\subsection{Industrial 4--20~mA and 0--10~V chains}
For the board dedicated to the 4--20~mA and 0--10~V industrial standards, a complete metrological bench characterization (absolute precision, long-term stability, and thermal drift) has not yet been carried out; instead, functional validation of scales, correct acquisition, and compatibility with real devices under operating conditions was performed.
Preliminary verification used precision resistors and a bench voltage source to check correspondence between expected values and ADC codes along significant portions of the scale and to validate the absence of saturation or discontinuities in the acquisition chain; in this phase, emphasis was placed on reproducibility and conversion consistency rather than on traceable absolute-error measurement.
Subsequently, validation of the 0--10~V channel was performed by connecting laboratory vacuum gauges (rough-vacuum and ultra-high-vacuum) equipped with standard 0--10~V analog outputs: qualitative comparison between the expected process evolution (monitored locally) and the acquired readout confirmed correct scale interpretation, measurement continuity, and the ability of the datalogger to follow slow variations and transients typical of pumping cycles.
For the 4--20~mA standard, realistic-condition verification was performed by acquiring temperature from an Optris \emph{CSlaser LT} infrared thermometer (CATCSLLTSFCK) used in vacuum with a 4--20~mA output: the readout was stable and consistent with expected transmitter behavior, confirming interface suitability for current-loop cabling and the effectiveness of time-series telemetry in promptly highlighting wiring anomalies, loop-power issues, or fault conditions.

\subsection{Power chain and system parameters}

A system-specific element, useful both during commissioning and
continuous operation, is monitoring internal operating parameters
through dedicated telemetry channels. In addition to experimental
measurements, the system continuously acquires the regulated analog
supply rails (\SI{+5}{V} and \SI{-5}{V}) and the regulated digital
supply rail (\SI{3.3}{V}) used by the module, together with the
internal board temperature.

\begin{figure}[t]
\centering
\includegraphics[width=0.90\textwidth]
{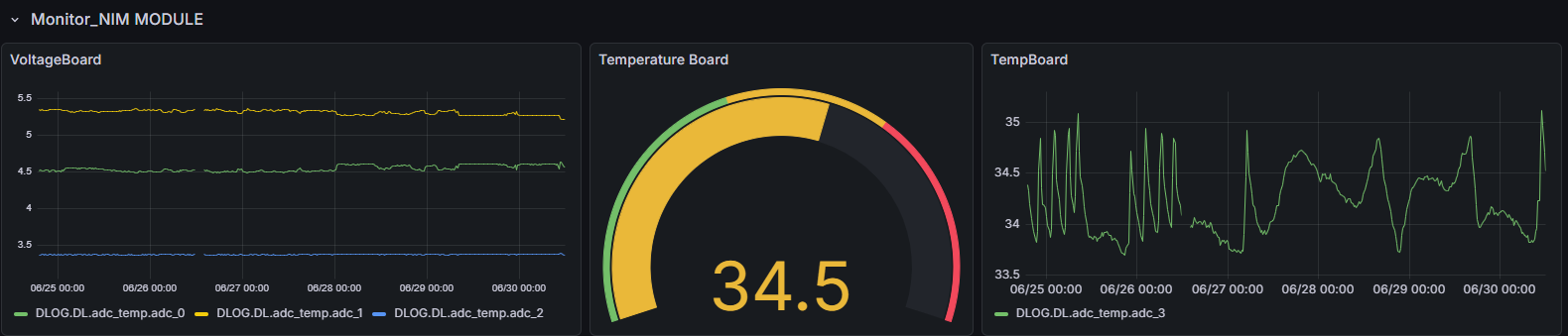}
\caption{
Example Grafana dashboard showing monitoring of the internal operating
parameters of the NIM module. The regulated analog supply rails
(\SI{+5}{V} and \SI{-5}{V}), the digital supply rail
(\SI{3.3}{V}), and the internal board temperature are continuously
acquired, archived, and visualized through the telemetry system.
}
\label{fig:power_monitoring}
\end{figure}

An example of the corresponding Grafana monitoring dashboard is shown
in figure~\ref{fig:power_monitoring}.

These parameters are stored in the time-series database and are
available through Grafana dashboards, allowing long-term supervision
of module health and direct correlation between operating conditions
and acquired measurements. This functionality provides significantly
more diagnostic information than simple front-panel status indicators
and facilitates troubleshooting during installation and operation.

\section{Conclusions}
We have described a modular datalogger and slow-control platform in NIM 2U format aimed at physics experiments requiring continuous and reliable monitoring of environmental and facility parameters.
The architecture integrates a base board with eight PT100/PT1000 channels and up to three extension boards, enabling rapid adaptation to heterogeneous sensor sets, including standard industrial inputs and a current-measurement channel in the \SI{10}{nA}--\SI{150}{\micro A} range.
A qualifying element of the project is the ``metrologically uniform'' approach: the systematic adoption of 16-bit SAR conversion and a common \SI{2.5}{V} reference across all boards establishes a shared and coherent conversion basis among subsystems, simplifying inter-channel comparisons, calibrations, and long-term maintenance.

On the software and operations side, integration with Graphite and Grafana provides a laboratory-ready telemetry and visualization pipeline in which parameters are stored as time series and made available via web dashboards.
A hierarchical, semantically stable metric naming scheme standardizes telemetry across subsystems and facilitates reuse of panels and dashboards across experimental campaigns, reducing reconfiguration effort while keeping the monitoring infrastructure unchanged.
The system is currently deployed in operation with positive feedback in terms of robustness and practical usefulness during shifts, and it remains an open platform for the development of additional extension boards should experiment-specific measurement or interfacing needs arise.

\acknowledgments

\bibliographystyle{JHEP}
\bibliography{references}
\end{document}